\documentclass[11pt]{article}
\usepackage{amsmath}
\usepackage{amsfonts}

\def\be{\begin{equation}}
\def\ee{\end{equation}}
\def\bea{\begin{eqnarray}}
\def\eea{\end{eqnarray}}

\def\hR{\hat{R}}
\def\hg{\hat{g}}
\def\hM{\hat{M}}
\def\hOmega{\hat{\Omega}}
\def\hT{\hat{T}}
\def\hLambda{\hat{\Lambda}}
\def\scri{\mathcal{I}}
\def\hR{\hat{R}}
\def\hh{\hat{h}}

\def\hC{\hat{C}}
\def\hA{\hat{A}}
\def\hB{\hat{B}}
\def\hu{\hat{u}}
\def\hg{\hat{g}}
\def\hM{\hat{M}}
\def\hm{\hat{m}}
\def\hOmega{\hat{\Omega}}
\def\hT{\hat{T}}
\def\hLambda{\hat{\Lambda}}
\def\ha{\hat{a}}
\def\hb{\hat{b}}
\def\cLambda{\check{\Lambda}}
\def\cR{\check{R}}
\def\ttau{\tilde{\tau}}
\def\cA{\check{A}}
\def\cB{\check{B}}
\def\cg{\check{g}}
\def\cm{\check{m}}
\def\cM{\check{M}}
\def\cT{\check{T}}
\def\ct{\check{t}}
\def\ck{\check{k}}
\def\ch{\check{h}}
\def\ca{\check{a}}
\def\cb{\check{b}}
\def\cC{\check{C}}
\def\cG{\check{G}}

\def\cLambda{\check{\Lambda}}

\def\cOmega{\check{\Omega}}
\def\cnabla{\check{\nabla}}
\def\crho{\check{\rho}}
\def\csigma{\check{\sigma}}
\def\cp{\check{p}}
\def\cT{\check{T}}

\begin{document}

\title{The equations of Conformal Cyclic Cosmology
}
\author{Paul Tod\\Mathematical Institute,\\University of Oxford,\\Oxford
OX2 6GG}

\maketitle
\begin{abstract}
I review the equations of Conformal Cyclic Cosmology given by Penrose \cite{p1}. Motivated by the example of FRW cosmologies, I suggest a slight modification to Penrose's prescription and show how this works out for 
Class A Bianchi cosmologies, and in general. 
\end{abstract}
\section{Introduction}
Penrose's Conformal Cyclic Cosmology (\cite{p1}, hereafter CCC) is a radical new cosmological model which assumes that the universe consists of a sequence of aeons, each of them a solution of the Einstein equations with sources 
and positive cosmological constant, each expanding from an initial Big Bang singularity to a space-like future null infinity $\scri$, the $\scri$ of one aeon being 
identified conformally with the Big Bang of the next. There is therefore a conformal metric common to consecutive aeons, which is assumed to be smooth, and three-surfaces $\Sigma$ which are both $\scri$ of 
one aeon and what we may call the \emph{bang surface} of the next. The surfaces $\Sigma$ are necessarily umbilic in a representative of the conformal metric. 

Other assumptions are made and we shall spell these out below by concentrating on two consecutive aeons. The purpose of this article is to set down equations for CCC and by 
consideration of examples seek some understanding of them.(The recent articles \cite{l1,n1} make different suggestions about equations for CCC.)

Thus we concentrate on two consecutive aeons, which we'll call past and present, so we have three manifolds-with-metric:
\begin{enumerate}
\item[$\bullet$] the previous aeon $(\hM,\hg_{ab})$; 
\item[$\bullet$] the present aeon $(\cM,\cg_{ab})$;
\item[$\bullet$] the conformal extension $(M,g_{ab})$ of both, so that there are conformal factors $\hOmega,\cOmega$ with
\be\label{g1}\hat{g}_{ab}=\hOmega^2g_{ab},\;\;\;\cg_{ab}=\cOmega^2g_{ab}.\ee
\end{enumerate}
We shall call $g_{ab}$ the \emph{bridging metric} and suppose that $M=\hM\cup\cM\cup\Sigma$ where $\Sigma$ is the common boundary:
\[\Sigma=\{\hOmega^{-1}=0\}=\{\cOmega=0\}.\]
Now necessarily $\Sigma$ is a singular surface for $\cM$, in that the curvature of $\cg$ is singular there, and we shall make enough assumptions below for $\Sigma$ to be $\scri$ for $\hM$. 
It also follows from (\ref{g1}) that the three Weyl tensors are equal:
\[\hC_{abc}^{\;\;\;\;\;d}=\cC_{abc}^{\;\;\;\;\;d}=C_{abc}^{\;\;\;\;\;d},\]
and then the Bianchi identity forces all three to vanish at $\Sigma$ (see Appendix B12 of \cite{p1}). This was one of the prime motivations for CCC, the desire to have Big Bang cosmological models 
with zero Weyl tensor at the bang.

Because of the conformal freedom, $g\rightarrow\Theta^2g$, $\Theta\neq 0$, in the choice of the bridging metric we can impose Penrose's \emph{reciprocal hypothesis}: 
\be\label{g2}\cOmega\hOmega=-1,\ee
and then necessarily, by (\ref{g1}), 
\be\label{g3}\cg_{ab}=\hOmega^{-4}\hg_{ab}\ee
so that the metric of the present aeon is determined by the metric of the previous aeon, provided we can give an algorithm to determine a unique $\hOmega$. This will be our main concern.

Penrose \cite{p1} assumes that the matter content in the previous aeon, at least near to $\scri$, is a radiation fluid together with a positive cosmological constant $\hLambda$. There will also be Maxwell fields: 
these are assumed not to be dynamically important but they will propagate through $\scri$ into the present aeon\footnote{where, for example, they might contribute to intergalactic magnetic fields.}. Now the matter 
content in $\cM$ near $\Sigma$, as determined through Einstein's equations by the 
Einstein tensor $\cG_{ab}$ of $\cg_{ab}$, will be fixed by $\hOmega$ and $\hg$. One no longer has freedom in choosing this matter content but it had better be roughly the same as in the previous aeon with, Penrose suggests, 
an extra field to represent dark matter.

In this scenario the previous aeon is straightforwardly defined as a solution of the Einstein equations with sources and $\Lambda>0$, and the Cauchy problem for this with data at $\scri$ is 
well-understood. Following Friedrich \cite{f1,f2} one has:
\begin{itemize}
\item[$\bullet$] for the vacuum Einstein equations with $\Lambda>0$, \cite{f1}, choose a Riemannian 3-metric $a_{ij}$ and a symmetric tensor $c_{ij}$ on $\Sigma=\scri$;
\item[$\bullet$] if $c_{ij}$ is trace-free and divergence-free w.r.t. the Levi-Civita connection of $a_{ij}$, then there is a solution of the Einstein equations with this data;
\item[$\bullet$] the data has gauge-freedom
\be\label{gf1}a_{ij}\rightarrow\tilde{a}_{ij}=\theta^2 a_{ij},\;\;\;c_{ij}\rightarrow\tilde{c}_{ij}=\theta^{-1} c_{ij},\ee
in that data related like this give the same solution.
\end{itemize}
(The free data $a_{ij}$ are the 3-metric of $\scri$, with a suitable choice of rescaling, and the free data 
$c_{ij}$ determine the normal derivative of the Weyl tensor at $\scri$; both appear in the Starobinski expansion (\ref{s1}) below.)

Furthermore:
\begin{itemize}
 \item[$\bullet$] the corresponding problem with any matter source described by a trace-free $T_{ab}$ and positive $\Lambda$ is formulated in \cite{f2}, and worked out in 
detail for Yang-Mills sources;
\item with a slightly different formalism, the corresponding result for a radiation fluid plus positive $\Lambda$ is proved in \cite{kl}.
\end{itemize}

Starting at the beginning of an aeon rather than at the end, there are some choices of matter model for which one can prove well-posed-ness of an initial value problem for the Einstein 
equations with data at a conformally-compactifiable initial singularity \cite{at1,at2, a, t1}, and solutions obtained this way will have a big bang at which 
the Weyl tensor is finite. However, one cannot expect that solutions to the 
Cauchy problem with data at $\scri$, as described above, will join on to these solutions with finite-Weyl tensor big bang. One way to see this, for example for radiation-fluid, 
is to observe that there 
is much less freely-specifiable data at the bang (just the 3-metric of $\Sigma$) than at $\scri$ (both $a_{ij}$ and $c_{ij}$). Furthermore, in this case, zero Weyl 
tensor at the bang forces the metric to be FRW \cite{at1}. Thus for CCC to work, the matter content after the bang cannot in general be as simple as a perfect fluid.

Rather than try to guess what it should be we adopt a different strategy, still following Penrose \cite{p1}: choose the desired matter model in $\hM$; devise a unique prescription for $\hOmega$; 
use the conformal rescaling to define the Einstein tensor $\cG_{ab}$ in $\cM$; and seek to interpret it.

There is an element of risk in this strategy in that the Einstein tensor after the bang may not be interpretable in terms of reasonable matter at all. The problem of finding a unique prescription for 
$\hOmega$ is taken up in Section 4, following as motivation an account of the FRW metrics in CCC in Section 3, and a review of the Starobinsky expansion of the metric near $\scri$ for positive $\Lambda$ in Section 2. The 
prescription for $\hOmega$ that I propose differs in detail from that suggested by Penrose in \cite{p1}. Then we 
consider the matter after the bang following the new strategy in Section 5, and we apply the new strategy to two classes of cosmological models in Sections 6 and 7. The results that we obtain are satisfactory to the first two 
orders in a post-bang time-coordinate.

\section{Starobinsky expansions}
Starobinsky  in a study of the asymptotics of metrics with positive $\Lambda$, \cite{s1},  considered metrics written in the form
\be\label{s1}g=dt^2-e^{2Ht}(a_{ij}+e^{-2Ht}b_{ij}+e^{-3Ht}c_{ij}+\ldots)dx^idx^j,\ee
where $\Lambda=3H^2$, $H>0$, and the spatial tensors $a_{ij},b_{ij},\ldots$ are time independent. This kind of expansion resembles the expansion of the Poincar\'e metric in the \emph{ambient metric construction} 
of Riemannian geometers \cite{fg}. To 
draw out the similarity to the present case, 
one can regard $r=e^{-Ht}$ as a defining function for $\scri$ and (\ref{s1}) as an expansion in the defining function, but there is a difference in that for the ambient metric construction the spatial metric should be 
analytic in $r^2$, which would forbid the term $c_{ij}$\footnote{which here is extra data; I am grateful to Pawel Nurowski for discussion of this point.}.

Starobinsky \cite{s1} argued that the general solution of the Einstein equations with fluid matter sources had an expansion like (\ref{s1}). 
Rendall \cite{r1} added a measure of rigour to the results of \cite{s1}, showing that there exist unique formal expansions of this form for large classes of solutions of the 
Einstein equations with perfect fluid sources, and that the vacuum solutions and solutions with Yang-Mills sources proved to exist by Friedrich all have expansions of this form (so 
(\ref{s1}) has `full asymptotics'). It is therefore reasonable to regard 
(\ref{s1}) as a general form of solution, at least near to $\scri$.

It is worth noting the existence of gauge freedom in an expansion like (\ref{s1}): the coordinate system is constructed by choosing a (late time) space-like surface to be $t=t_0=$ constant; then $t$ is proper time along the orthogonal congruence to this surface and the space-coordinates $x^i$ are co-moving. Thus one can make a new choice of the initial surface to produce a change
\[t\rightarrow\tilde{t}=t-\chi(x^i)\]
which must be accompanied by a redefinition of the comoving space coordinates. This transformation entails 
\[a_{ij}\rightarrow\tilde{a}_{ij}=e^{2H\chi} a_{ij},\;\;\;c_{ij}\rightarrow\tilde{c}_{ij}=e^{-H\chi} c_{ij},\]
which is (\ref{gf1}) with $\theta=e^{H\chi}$.

\bigskip

Our strategy will be to take $\hg_{ab}$ in the Starobinski form and seek a unique prescription for $\hOmega$. Following \cite{s1} we note that for the Einstein equations with energy momentum tensor $T_{ab}=O(e^{-3Ht})$, 
necessarily $b_{ij}$ 
is related to $a_{ij}$ by
\be\label{s2}
H^2b_{ij}=-(R^{(a)}_{ij}-\frac14R^{(a)}a_{ij}),\ee
where $R^{(a)}_{ij}$ and $R^{(a)}$ are the Ricci tensor and Ricci scalar of the metric $a_{ij}$. (Here and throughout we use conventions as in \cite{pr}.)

\section{The example of FRW}
The FRW metric sits very well with CCC as we shall see. Then guided by that calculation, we'll extend the method to progressively wider classes of cosmologies. 

Consider the FRW metric
\be\label{fr1}g=dt^2-R(t)^2d\sigma_k^2\ee
where $d\sigma_k^2$ with $k=-1,0,1$ is the usual constant curvature metric. If we assume a radiation fluid source,  then the conservation equation can be integrated to give the density as $\rho=mR^{-4}$ for a constant of integration $m$. The Einstein equations, with cosmological constant $\Lambda$ now reduce to the Friedmann equation, which is
\be\label{fr2}
\left(\frac{dR}{d\tau}\right)^2=\frac{m}{3}-kR^2+\frac{\Lambda}{3} R^4,\ee
where we have written this in terms of \emph{conformal time} $\tau$ defined by integrating $d\tau=dt/R$.

We'll suppose that the metric in the past aeon took this form, and therefore put hats on everything:  $\hg$, $\hR$, $\hat{t}$, $\hat{\rho}$, $\hLambda$, ... 

Now an obvious candidate for $\hOmega$ is a multiple of the scale factor: $\hOmega=c_1\hR$ 
for some constant $c_1$ to be fixed. With this choice
\be\label{fr3}\cg=\hOmega^{-4}\hg=d\ct^2-\cR(\ct)^2d\sigma_k^2,\ee
with 
\[\cR=-c_1^{-2}\hR^{-1},\;\;\;d\tau=d\hat{t}/\hR=d\ct/\cR.\](The minus sign in the first equation here has the same origin as the one in (\ref{g2}): $\cR$ goes through zero at the bang and so is negative in $\hM$.)
With this choice,  $\cg$ is again in the FRW form, and  with the choice $c_1=(\hLambda/\hm)^{1/4}$ the Friedmann equation transforms from a hatted to a checked version with\[\cm=\hm,\;\;\;\;\cLambda=\hLambda;\]
so the two aeons are \emph{diffeomorphic} i.e they are the same solution of the EFEs.

It will be helpful below to note here that, introducing $\phi=\hOmega^{-1}$, the Friedmann equation implies
\be\label{fr5}\hat{\Box}\phi+2H^2\phi=kc_1^2\phi^3,\ee
from which it follows that the bridging metric has scalar curvature $6kc_1^2$, and the solution for $\phi$ takes the form 
\be\label{fr6}\phi=\phi_1e^{-H\hat{t}}+O(e^{-3H\hat{t}})\ee with constant $\phi_1$.

\medskip

The FRW metric is of course very special, being conformally flat. However we may take this case, or more precisely 
these three cases as $k$ varies, as the paradigm and seek to stay close to them even when there is some Weyl curvature.

\section{Finding a unique $\hOmega$}
Suppose now we have a more general metric $\hg$ in the previous aeon which we take to be in Starobinski form (\ref{s1}), but with hats. We may expand $\hOmega$ in the corresponding way as
\be\label{fi1}\phi:=\hOmega^{-1}=e^{-H\hat{t}}\phi_1+e^{-2H\hat{t}}\phi_2+
e^{-3H\hat{t}}\phi_3+\ldots,\ee
and then the metric of $\scri$ will be $\phi_1^2\ha_{ij}$.

We can find an equation for $\phi$ by noting, in line with (\ref{fr5}), that the scalar curvature $s$ of the bridging metric $g$ satisfies
\be\label{fi2}\hat{\Box}\phi +2H^2\phi=\frac{1}{6}s\phi^3. \ee
Here we've used the fact that the scalar curvature of the previous aeon is $4\hLambda=12H^2$. With $s=12H^2$, (\ref{fi2}) is Penrose's \emph{phantom field equation} \cite{p1}, 
but we can leave $s$ as a constant to be chosen later and proceed to solve this equation.

Substituting (\ref{fi1}) into (\ref{fi2}) and solving term by term we find that  $\phi_1$ and $\phi_2$ are freely specifiable, with  subsequent $\phi_n$ then determined. 
The example of FRW, (\ref{fr6}), suggests choosing $\phi_2=0$ (which is Penrose's \emph{Delayed Rest Mass Hypothesis} \cite{p1}). To specify $\phi_1$, again following the example of 
FRW, we can seek to give the metric of $\scri$, which is $\phi_1^2\ha_{ij}$, constant scalar curvature. Doing this is essentially solving 
the (three-dimensional) Yamabe problem (see e.g. \cite{yam} or \cite{ma}) and what we can say is that, for a compact $\scri$,  $\phi_1$ should be chosen to minimise the Yamabe functional for $\ha_{ij}$. 
This prescription may not work for a noncompact $\scri$ but at least for a compact $\scri$ we have a prescription for a unique $\phi$ up to multiplication by a constant and hence a unique $\hOmega$ up to a constant. Henceforth, 
we shall assume that this choice has been made, so that $\ha_{ij}$ has constant scalar curvature and so $\phi_1$ is a constant. For $\phi_3$ we obtain:
\be\label{fi5}2H^2\phi_3=\frac16s\phi_1^3-H^2\phi_1\hb=\frac{1}{12}\phi_1R^{(\ha)},\ee
where we've used (\ref{s2}) to eliminate $\hb$, and $R^{(\ha)}$ is the scalar curvature of the metric $\ha_{ij}$.

It still remains to fix $s$ in (\ref{fi2}) and again we'll be guided by the example of FRW, where $s=6kc_1^2$ which (with the conventions of \cite{pr}) 
is $-s^{\mathcal{I}}$ with  $s^{\mathcal{I}}$ the scalar curvature of the metric of $\scri$ .
These choices fix $\hOmega$ or equivalently $\phi$ up to a single constant $\phi_1$
which we may hope to fix by demanding $\cLambda=\hLambda$. This remaining constant could be taken to be the (now constant) scalar curvature of $\scri$ but we shall instead choose it to be 
\be\label{fi4}
a_1:=-\mbox{Lim}\left(\frac{d\phi}{d\tau}\right),\ee
where the limit is taken at $\scri$ (this limit is $H\phi_1$).
 
\section{The matter content after the bang}
The rescaling formula for the Ricci tensor with $\cg=\hOmega^{-4}\hg$, and conventions as in \cite{pr}, is
\be\label{res1}\hR_{ab}=\cR_{ab}+2\cnabla_a\Upsilon_b-2\Upsilon_a\Upsilon_b+\cg_{ab}\cg^{ef}(\cnabla_e\Upsilon_f+2\Upsilon_e\Upsilon_f)\ee
with $\Upsilon_a=2\partial_a\log\hOmega$. The Einstein equations in $\hM$ are
\[\hR_{ab}=-\kappa\hT_{ab}+\hLambda\hg_{ab},\]
where
\[\hT_{ab}=\frac13\hat{\rho}(4\hu_a\hu_b-\hg_{ab}).\]

To preserve the conservation equation we define
 \[\cT_{ab}=\hOmega^4\hT_{ab}=\phi^{-4}\hT_{ab} \]
 though this shouldn't be regarded as accounting for all the matter content after the bang. Then solving (\ref{res1}) for  $\cG_{ab}$
we find
\be\label{res2}\cG_{ab}=-\kappa\phi^{4}\cT_{ab}+\frac{4}{\phi}\cnabla_a\cnabla_b\phi+\frac{4}{\phi^2}\phi_a\phi_b+\left(8\frac{|\cnabla\phi|^2}{\phi^2}-4\frac{\check{\Box}\phi}{\phi}-\frac{\hLambda}{\phi^4}\right)\cg_{ab},\ee
so that there is a contribution to the post-bang matter content from the scalar field $\phi$. Since we have no freedom to change the expression (\ref{res2}) we have to ask if it gives a sensible answer. 
We approach this question first via an example.

\section{An example: Class A Bianchi types}
In this section we apply the procedure to spatially-homogeneous cosmological models with the symmetry of class A Bianchi types. These generically have nonzero Weyl curvature. 
Before restricting to CCC it is worth noting a few facts about the Bianchi metrics. Assume the metric to be
\be\label{b1}g=dt^2-R^2(e^{2\alpha}\sigma_1^2 +e^{2\beta}\sigma_2^2+e^{2\gamma}\sigma_3^2),\ee
in terms of four functions of time,   
$(R(t),\alpha(t),\beta(t),\gamma(t))$, with $\alpha+\beta+\gamma=0$, and the usual set of invariant one-forms $\sigma_i$, which are taken to satisfy
\be\label{b3}
d\sigma_1=n_1\sigma_2\wedge\sigma_3,\ee
and cyclic permutations, where each $n_i$ is $\pm 1$ or $0$.

In terms of the orthonormal tetrad $(\theta^0, \theta^1, \theta^2,\theta^3)=(dt,Re^\alpha\sigma_1,Re^\beta\sigma_2,Re^\gamma\sigma_3)$ the components of the Einstein tensor (still with conventions as in \cite{pr}) are
\begin{eqnarray}\label{b5}
 G_{00}&=&-\left(3\frac{\dot{R}^2}{R^2}+\dot{\alpha}\dot{\beta}+\dot{\beta}\dot{\gamma}+\dot{\gamma}\dot{\alpha} +AB+BC+CA\right)\\
 G_{11}&=&-\left(-2\frac{\ddot{R}}{R}-\frac{\dot{R}^2}{R^2}+3\frac{\dot{R}\dot{\alpha}}{R}+\ddot{\alpha}-\dot{\alpha}^2+\dot{\beta}\dot{\gamma}-AB+BC-CA\right)\nonumber
\end{eqnarray}
with the overdot being $d/dt$, cyclic permutations giving $G_{22},G_{33}$, and all other components zero. Also here
\[A=\frac{1}{2R}(-n_1e^{2\alpha}+n_2e^{2\beta}+n_3e^{2\gamma}),\]
and cyclic permutations for $B,C$.

We can also note the components of the Weyl tensor for the metric (\ref{b1}) in the orthonormal basis. These are
\bea\label{w1}
E_{11}&=&-\frac{\dot{R}\dot\alpha}{R} +\frac13(2\dot\beta\dot\gamma-\dot\gamma\dot\alpha-\dot\alpha\dot\beta-4BC+2CA+2AB), \\\label{w2}
B_{11}&=&\dot\alpha(B+C)-\dot\beta C-\dot\gamma B,
\eea
with all other nonzero components obtained by symmetry.

With a metric of this form in the previous aeon, $\scri$ will be at $\hat{t}=\infty$, The leading term $a_{ij}$ in the Starobinsky expansion (\ref{s1}) 
will have constant scalar curvature so the leading term $\phi_1$ in $\phi$ in (\ref{fi1}) will be constant. This forces $\phi$ and hence $\hOmega$ to be functions only of $\hat{t}$ 
and therefore $\cg$ has the same symmetry as $\hg$. If $\hg,\cg$ are both of the form of (\ref{b1}) with respectively hats and checks on all quantities then
\[\cR=-\phi^2\hR,\;\;d\check{t}=-\phi^2d\hat{t},\;\;\check{\alpha}=\hat{\alpha}\mbox{   etc.}\]
From the expression for $A,B,C$ it 
also follows that $\check{A}=\phi^{-2}\hat{A}$, and permutations of this.

We shall suppose that $\hM$ contains radiation fluid plus $\hLambda$ so that the Einstein equations in the previous aeon become
\[\hat{G}_{00}=-(\hat{\rho}+\hLambda),\;\;\hat{G}_{11}=\hat{G}_{22}=\hat{G}_{33}=-(\frac13\hat{\rho}-\hLambda).\]
From the symmetry, it is inevitable that the Einstein tensor of $\cg$ is also diagonal but it won't have the perfect-fluid form. We set
\be\label{b7}\check{G}_{00}=-(\check{\rho}+\cLambda),\;\;\check{G}_{11}=-(\check{p}_1-\cLambda),\;\;\check{G}_{22}=-(\check{p}_2-\cLambda),\;\;\check{G}_{33}=-(\check{p}_3-\cLambda),\ee
in terms of \emph{principal pressures} $\check{p}_i$, and then we can find these diagonal 
entries as power series in $\check{t}$, the proper-time coordinate in $\cM$. This is most easily accomplished by calculating first in conformal time $\tau$ where
\[d\tau=d\hat{t}/\hR=d\check{t}/\cR,\]
and this was done in \cite{t1}, from which we quote. The (00) Einstein equation in $\hM$ forces $\hR$ to blow up at finite $\tau$. Suppose this is at $\tau=\tau_F$ and set $\tilde\tau=\tau_F-\tau$, then
\begin{eqnarray}\label{b9}
\hR&=&\frac{1}{H\ttau}(1+a\ttau^2+O(\ttau^3))\\\nonumber
\alpha&=&\alpha_0+\alpha_1\ttau^2+O(\ttau^3)\\\nonumber
\beta&=&\beta_0+\beta_1\ttau^2+O(\ttau^3)\\\nonumber
\gamma&=&\gamma_0+\gamma_1\ttau^2+O(\ttau^3)\\\nonumber
\phi&=&a_1\ttau+a_3\ttau^3+O(\ttau^4)\nonumber
\end{eqnarray}
where $a=\mbox{Lim }\hR^2(AB+BC+CA)/18$, and the limit is onto $\scri$; $a_1\neq 0$ is a free constant in $\phi$; $\alpha_0,\beta_0,\gamma_0$ are the values of $\alpha,\beta,\gamma$ at $\scri$ and, with $a_1$, determine 
the metric of $\scri$ which is free data; $\alpha_1=\mbox{Lim}\frac{2}{3}\hR^2(2BC-AB-AC)$ with $\beta_1,\gamma_1$ obtained by cyclic permutation; and $a_3$ is obtained by solving (\ref{fi2}) as
\[a_3=-4aa_1+\frac{a_1^3s}{12H^2}.\]
We can simplify $a_3$ with the aid of the metric of $\scri$, which is 
 \be\label{b2}\frac{a_1^2}{H^2}(e^{2\alpha_0}\sigma_1^2 +e^{2\beta_0}\sigma_2^2+e^{2\gamma_0}\sigma_3^2).\ee
The scalar curvature of (\ref{b2}) is $s^{\mathcal{I}}=-\frac{36H^2a}{a_1^2}$ and our prescription above was to take $s=-s^{\mathcal{I}}$, in which case $a_3=-aa_1$. Note also from (\ref{b2}) that the Ricci tensor of $\scri$ 
is diagonal with components determined by
\be\label{b4}
R^{\scri}_{11}-\frac{1}{3} s^{\scri}=-\frac{H^2}{a_1^2}\alpha_1,\ee
and cyclic permutations of this.
 
\medskip

For the Weyl tensor components we find from (\ref{w2}) that $B_{ij}=O(\ttau^3)$ while from (\ref{w1}) $E_{ij}=O(\ttau^2)$ (recall that the Weyl tensor is expected to vanish at $\scri$, 
and these components are conformally invariant, so that indeed it does). 

For the matter content after the bang we start with (\ref{b7}) and (\ref{b5}):
\[\check{\rho}+\cLambda=-\check{G}_{00}=3\frac{\check{R}^2_\tau}{\check{R}^4}+\frac{1}{\cR^2}(\alpha_\tau\beta_\tau+\beta_\tau\gamma_\tau+\gamma_\tau\alpha_\tau) +\cA\cB+\cB\cC+\cC\cA\]
\[=\frac{3}{\phi^8\hR^4}\left(\frac{d(\phi^2\hR)}{d\tau}\right)^2+\frac{1}{\phi^4\hR^2}(\alpha_\tau\beta_\tau+\beta_\tau\gamma_\tau+\gamma_\tau\alpha_\tau) +\frac{1}{\phi^4}(\hA\hB+\hB\hC+\hC\hA)\]
\[=\frac{1}{\phi^4}(\hat{\rho}+\hLambda)+\frac{12}{\phi^6\hR^3}\frac{d\phi}{d\tau}\frac{d(\phi\hR)}{d\tau}\]
\[=\frac{3H^2}{a_1^4\ttau^4}+\frac{12H^2a}{a_1^4\ttau^2}+O(1),\]
using the expansions of (\ref{b9}) in the last line.

For a principal pressure we have
\[\check{p}_1-\cLambda=-\check{G}_{11}=-2\frac{\ddot{R}}{R}-\frac{\dot{R}^2}{R^2}+3\frac{\dot{R}\dot{\alpha}}{R}+\ddot{\alpha}-\dot{\alpha}^2+\dot{\beta}\dot{\gamma}-\cA\cB+\cB\cC-\cC\cA\]
\[=-\frac{2}{\cR^2}\frac{d}{d\tau}\left(\frac{1}{\cR}\frac{d\cR}{d\tau}\right)-\frac{3}{\cR^4}\left(\frac{d\cR}{d\tau}\right)^2+\frac{3}{\cR^3}\frac{d\cR}{d\tau}\frac{d\alpha}{d\tau}\]\[+
 \frac{1}{\cR}\frac{d}{d\tau}\left(\frac{1}{\cR}\frac{d\alpha}{d\tau}\right)+\frac{1}{\cR^2}\left(\frac{d\beta}{d\tau}\frac{d\gamma}{d\tau}-(\frac{d\alpha}{d\tau})^2\right)+\frac{1}{\phi^4}(-\hA\hB+\hB\hC-\hC\hA).\]
 Put $\cR=-\phi^2\hR$ and simplify to find
\[=\frac{1}{\phi^4}(\frac13\hat{\rho}-\hLambda)+\frac{4}{\phi^5\hR^2}\frac{d\alpha}{d\tau}\frac{d\phi}{d\tau}-\frac{4}{\phi^6\hR^3}\frac{d\phi}{d\tau}\frac{d(\phi\hR)}{d\tau}
-\frac{4}{\phi^4\hR^2}\frac{d}{d\tau}\left(\frac{1}{\phi}\frac{d\phi}{d\tau}\right),\]
\[=\frac{H^2}{a_1^4\ttau^4}+\frac{4H^2(a+2\alpha_1)}{a_1^4\ttau^2}+O(1).\]
It is natural to write quantities after the bang in terms of proper time $\ct$ rather than conformal time so that we need an expression for $\ct$ in terms of $\ttau$ which we obtain by solving
\[d\ct=\cR d\tau=-\phi^2\hR d\tau.\]
We find
\[\ct=\frac{a_1^2}{2H}\ttau^2-\frac{aa_1^2}{4H}\ttau^4+\mbox{h.o.}\]
choosing a common origin for $\ct$ and $\ttau$, which inverts to
\[\ttau^2=\frac{2H}{a_1^2}\ct+\frac{2H^2a}{a_1^4}\ct^2+\mbox{h.o.}.\]
In terms of $\ct$ we find
\bea\label{b10}
\crho&=&\frac{3}{4\ct^2}+\frac{9aH}{2a_1^2\ct}+O(1)\\
\check{p}_1&=&\frac{1}{4\ct^2}+\frac{H}{2a_1^2\ct}(3a+8\alpha_1)+O(1).\nonumber
\eea
This is our final result in this section: to leading order, which is $O(\ct^{-2})$, the matter after the bang is a radiation perfect fluid with isotropic pressures equal to one-third of density; to next order the pressure $\cp_1$ differs 
from one-third of density by a multiple of $\alpha_1$ which in turn is a multiple of $R^{\scri}_{11}-\frac13 s^{\scri}$; the sum of the pressures is still equal to the density at this order, so that the 
energy-momentum tensor is still trace-free; at the next order, the sum of the pressures will cease to equal the density. 

These are satisfactory answers, to these orders. The next step is to see what happens without the assumption of spatial homogeneity.

\section{After the bang: the general case}
Now we repeat the calculation of the previous section but without the assumption of spatial homogeneity, so the metric in the previous aeon is of the Starobinsky form (\ref{s1}) but with hats. 
Then the space-time metric after the bang, from (\ref{g3}), can be written
\[\cg=\phi^4\hg=(e^{-H\hat{t}}\phi_1+e^{-3H\hat{t}}\phi_3+O(e^{-4H\hat{t}}))^4(d\hat{t}^2-\hh_{ij}dx^idx^j)\]
with constant $\phi_1$ and $\phi_3$, and with $\hh_{ij}$ read off from (\ref{s1}). Introduce a post-bang time coordinate $\cT$ by
\[d\cT=(e^{-H\hat{t}}\phi_1+e^{-3H\hat{t}}\phi_3)^2d\hat{t},\]
(note that this differs from $\ct$ in the previous section since $\phi$ will generally be dependent on $x^i$). Integrate for $\cT$, then
\[\cg=Vd\cT^2-\ch_{ij}dx^idx^j,\]
with
\[V=\left(\frac{e^{-H\hat{t}}\phi_1+e^{-3H\hat{t}}\phi_3+O(e^{-4H\hat{t}})}{e^{-H\hat{t}}\phi_1+e^{-3H\hat{t}}\phi_3}\right)^4=1+O(\cT^{3/2}),\]
and
\[\ch_{ij}=\cT\ca_{ij}+\cT^2\cb_{ij}+O(\cT^{5/2})\]
with
\bea
\ca_{ij}&=&2H\phi_1^2\ha_{ij}\\\label{bb2}
\cb_{ij}&=&4H^2\hb_{ij}+12H^2\frac{\phi_3}{\phi_1}\ha_{ij}.\label{bb3}
\eea
From (\ref{bb3}), (\ref{s2}) and (\ref{fi5}) we deduce
\be\label{bb5}\cb=\frac12R^{(\ca)}.\ee
We want to calculate the post-bang Einstein tensor in order to find the post-bang matter content. 

The inverse metric after the bang is
\[\cg^{-1}=V^{-1}\left(\frac{\partial}{\partial\cT}\right)^2-\ck^{ij}\partial_i\partial_j\]
with
\[\ck^{ij}=\frac{1}{\cT}\ca^{ij}-\cb^{ij}+O(\cT^{3/2}),\]
where indices are raised and lowered with $\ca^{ij}$ and $\ca_{ij}$ which are inverses. Then a straightforward but tedious calculation gives coordinate components of the Einstein tensor for $\cg$:
\[\cG_{00}=-\frac{3}{4\cT^2}+\frac{1}{2\cT}(R^{(\ca)}-\cb)+O(\cT^{-1/2}),\]
\[\cG_{0i}=O(1),\]
\[\cG_{ij}=-\frac{1}{4\cT}\ca_{ij}+R^{(\ca)}_{ij}-\frac12 R^{(\ca)}\ca_{ij}-\cb_{ij}+\frac34\cb\ca_{ij}+O(\cT^{1/2}).\]
We wish to interpret the Einstein tensor $\cG_{ab}$ in terms of matter after the bang. We set
\be\label{bb1}\crho:=-\cG_{00}=\frac{3}{4\cT^2}-\frac{1}{4\cT}R^{(\ca)}+O(\cT^{-1/2}),\ee
to define $\crho$ (and we've used (\ref{bb5})) and we set
\be\label{bb6}\cG_{ij}=-\frac13\crho\ch_{ij}-\csigma_{ij}+O(\cT^{1/2}),\ee
(where there is no assumption yet that $\csigma_{ij}$ is trace-free). Then with the aid of identities already obtained we find
\be\label{bb7}\csigma_{ij}=-4(R^{(\ca)}_{ij}-\frac13R^{(\ca)}\ca_{ij})\sim-4(R^{(\ha)}_{ij}-\frac13R^{(\ha)}\ha_{ij}),\ee
which is in fact trace-free and should be compared with the result (\ref{b4}) of the previous section.

Thus what we found in the previous section does carry over to the general case: to leading order, which is $O(\cT^{-2})$, the matter after the bang is radiation perfect fluid with 
isotropic pressures equal to one-third of density; to next order, $O(\cT^{-1})$, the spatial part of the energy-momentum tensor is a term equal to one-third of density times spatial metric plus a trace-free term which 
is a multiple of the trace-free Ricci tensor of $\scri$.


\end{document}